%Paper: hep-ph/9307303
%From: ereidell@marie.mit.edu (Evan Reidell)
%Date: Mon, 19 Jul 1993 23:22:12 -0400

%%% ctp2201.tex
\magnification=1200
\vsize=7.5in
\hsize=5.6in
\tolerance 500
\baselineskip 12pt plus 1pt minus 1pt
\pageno=0
\centerline{\bf INTERPOLATING THE NUCLEON'S SPIN-DEPENDENT}
\centerline{{\bf SUM RULES AT HIGH AND LOW ENERGIES}
\footnote{*}{This work is supported in part by funds provided by the
U. S. Department of Energy (D.O.E.) under contract  \#DE-AC02-76ER03069.}}
\vskip 24pt
\centerline{Xiangdong Ji}
\vskip 12pt
\centerline{\it Center for Theoretical Physics}
\centerline{\it Laboratory for Nuclear Science}
\centerline{\it and Department of Physics}
\centerline{\it Massachusetts Institute of Technology}
\centerline{\it Cambridge, Massachusetts\ \ 02139\ \ \ U.S.A.}
\vskip 1.5in
\centerline{Submitted to: {\it Phys. Lett. B}}
\vfill
\baselineskip 24pt plus 2pt minus 2pt
\centerline{\bf ABSTRACT}
\medskip
I discuss a few interpolating sum rules for spin structure functions of the
nucleon.  Using the concept of duality, I argue that the $G_1$ sum rule,
including the elastic contribution, is useful for learning higher twist matrix
elements of the nucleon.
\vfill
\line{CTP\#2201 \hfil April 1993}
\eject

Recently, there has been some interest in measuring the nucleon's
spin-dependent structure functions at moderate and low virtual-photon mass,
$Q^2$, with electron scattering.$^1$ This is motivated by the observation that
the first moment of the proton's spin structure function $G_1(x,Q^2)$ has been
measured by EMC at an averaged $Q^2$ = 10 GeV$^2$ and is positive,$^2$
however, the moment seems to become negative at the real photon point
according to the celebrated Drell-Hearn-Gerasimov sum rule.$^{3,4}$ Thus, an
interesting question that arises immediately is how the moment changes with
$Q^2$ and what physics causes such a change.

Unfortunately, there are some controversies in the
literature about defining a $G_1$
sum rule for all $Q^2$ and physical significance
of its $Q^2$ variation. Moreover, for
longitudinally-polarized
virtual-photon scattering, the nucleon's other
spin-dependent structure function, $G_2$,
also contributes at low and moderate $Q^2$.
The purpose of this paper is to
clarify some of these issues.

To start, let me quote the standard definition
of the nucleon's spin-dependent structure functions.
In inclusive electron or photon scattering, one measures
the nucleon tensor,
$$      W_{\mu\nu} = \int d^4\xi e^{i\xi\cdot q}
         \langle PS|J_{\mu}(\xi) J_{\nu}(0)|PS\rangle,   \eqno(1) $$
where $J_\mu$ is the electromagnetic current of the
nucleon, $|PS\rangle$ is the nucleon state with
momentum $P$ and polarization $S$, and $q$ is the virtual-photon
four momentum. The
spin-dependent part of the tensor is antisymmetric
in $\mu$ and $\nu$
and can be characterized by
the following two spin structure
functions ($\epsilon^{0123} = 1$),
$$     W_{\mu\nu} = -i\epsilon_{\mu\nu\rho\sigma}
            q^{\rho}\Big[{1\over M^2} G_1 S^\sigma +
             {1\over M^4} G_2 ( S^\sigma \nu -
              P^\sigma (S\cdot q))\Big],  \eqno(2)  $$
where $\nu= P\cdot q$.
In the Bjorken limit, one can define the corresponding
scaling functions,
$$ \eqalign{       g_1(x, Q^2) & = {\nu\over M^2} G_1(\nu, Q^2),   \cr
        g_2(x, Q^2) & = \left( {\nu\over M^2} \right)^2 G_2 (\nu, Q^2). }
   \eqno(3)   $$

Let me first consider the structure function $G_1$.
Following ref. 4, I define a $Q^2$-dependent integral,
$$ \eqalign{         I_1(Q^2) & = \int^{\infty}_{Q^2/2}
              {d\nu \over \nu} G_1(\nu, Q^2),   \cr
             & = {2M^2\over Q^2}\int^1_0 g_1(x, Q^2) dx,} \eqno(4) $$
where the lower limit in the integration includes
the elastic contribution.
The EMC date shows,$^2$
$$         \int^1_0 dx g_1(x)|_{\bar Q^2=10 {\rm GeV^2}} = 0.126
          \pm 0.010 \pm 0.015,  \eqno(5) $$
for the proton.
On the other hand, I will show below that
$$          I_1(Q^2) = {M^2\over Q^2} F_1(Q^2)(F_1(Q^2) + F_2(Q^2))
            - {\kappa^2\over 4}, \eqno(6) $$
as $Q^2\rightarrow 0$. Here $F_1$ and $F_2$ are
the Dirac and Pauli form factors of the nucleon, and
$\kappa = F_2(0)$ is the anomalous magnetic moment.
The first term in eq. (6)
comes from the elastic scattering and
the second term represents the inelastic
contributions summed by the Drell-Hearn-Gerasimov
sum rule.

The elastic contribution to $G_1$ is well-known
theoretically and has been measured experimentally
at one $Q^2$.$^5$ Its presence in $I_1(Q^2)$
at low $Q^2$ can be seen in the following way. Consider the $G_1$
dispersion integral,
$$        S_1(\nu, Q^2) = 4 \int^\infty_{Q^2/2}
         {\nu' d\nu' \over \nu'^2-\nu^2}
              G_1(\nu', Q^2),   \eqno(7)  $$
where $S_1(\nu, Q^2)$ is the corresponding spin-dependent
forward Compton amplitude. A simple calculation
of the nucleon pole diagram yields, in the soft photon limit,
$$      S_1(\nu, Q^2) = -2(F_1+F_2)F_1M^2\Big[
           {1\over 2\nu-Q^2} - {1\over 2\nu + Q^2} \Big]
           - F_2^2.    \eqno(8)  $$
Two different soft photon limits can be taken in eq. (7)
to obtain $G_1$ sum rules. If one takes $Q^2 \rightarrow 0$
first, $S_1(\nu, 0)= -\kappa^2$ at small $\nu$
and the elastic contribution to the $G_1$ integral
vanishes, then eq. (7) is just the Drell-Hearn-Gerasimov
sum rule. On the other hand, if one takes
$\nu\rightarrow 0$ first, $S_1(0, Q^2) = 4I_1(Q^2)$, where
$I_1$ is given by eq. (6), the elastic contribution
remains in the $G_1$ integral in eq. (7). Obviously,
these two limiting processes do not produce
any conflicting results. However, to obtain
the sum rule (4) at low $Q^2$, one should take
the second limit.

One might insist that the elastic contribution
vanishes rapidly at high $Q^2$ and thus it is
equally interesting to consider the elastic-subtracted
sum rule $\bar I_1(Q^2)$ to interpolate
the Drell-Hearn-Gerasimov sum rule and the EMC
result.$^4$ However, I argue below that $I_1(Q^2)$
has richer physical content and is more useful
in practice.
To demonstrate this, imagine the nucleon
is a structureless spin-half particle, then at all $Q^2$,
$$      I_1(Q^2)  = {M^2\over Q^2}.  \eqno(9)    $$
This result can be explained in
terms of either pure elastic scattering or
deep-inelastic scattering on a simple structure.
Both languages dual each other in the entire range of
$Q^2$. For the nucleon with a non-trivial structure,
$I_1(Q^2)$ deviates from eq. (9) at high and
low $Q^2$ for different physics reasons. At high $Q^2$
where the deep-inelastic structure is of relevant, the
nucleon is a superposition of free point-like quarks
and the coefficient in eq. (9) is modified
by the quark helicity distribution probability,
$2\int g_1(x)$. At low $Q^2$, $I_1(Q^2)$ still has a $1/Q^2$
behavior since to a small mass virtual-photon,
the nucleon is a point-like particle. However, its coupling
with the photon is modified by its anomalous magnetic
moment. Then, an interesting question is
how to understand the $Q^2$ evolution of
$I(Q^2)Q^2/M^2$ from $1+\kappa$ at $Q^2=0$ to $2\int g_1(x)$ at
high $Q^2$?

As $Q^2$ changes from high to low, the interactions
between quarks in the nucleon become important, and $I(Q^2)$
acquires higher order terms in $1/Q^2$ expansion,
which are the higher twist effects.
On the other hand, when $Q^2$ increases
from 0, the nucleon resonance contribution to
$I(Q^2)$ starts to dominate. It may happen,
however, that there exists a region of $Q^2$,
presumably around 1 GeV$^2$, where $I(Q^2)$ can
be explained in both the deep-inelastic and
resonance physics languages. This duality phenomena
was first observed in the unpolarized
deep-inelastic scattering.$^6$
One useful consequence of duality is that it
allows to extract matrix elements of
higher-twist operators from data in
the resonance-dominated region.

The size of the duality region is difficult to
access. If in the region the elastic contribution
is not important, then it is a matter of free
choice to use either $I_1(Q^2)$ or $\bar I_1(Q^2)$
to extract higher twists. However, if the
region extends to small $Q^2$ where the
elastic contribution is dominant, the dual of
deep-inelastic physics clearly contains the
elastic contribution, as seen in a simple example
in eq. (9). Therefore, to learn about the higher twist
effects at small $Q^2$, the elastic contribution must
be included in the $G_1$ sum rule. This in fact is
well-known in the case of unpolarized scattering.$^7$
According to this observation, the method
used in refs. 4 and 8 of extracting higher
twists is unreliable. Recently, Unrau
and I have studied the twist-four contribution
and target mass corrections to $I_1(Q^2)$ in QCD and
the bag model,$^9$ extending and
improving the previous studies on this problem.$^{10}$

For the neutron, the elastic contribution
of the order of $1/Q^2$ to $I_1(Q^2)$ vanishes at $Q^2 =0$.
A constant term is generated from its anomalous
magnetic moment and charge radius, $\kappa(\kappa/4
-M^2<r^2>_{\rm cm}/6)$. Due to a remarkable numerical
coincidence, this combination is essentially zero.

No data exist so far on the other spin-dependent structure
function $G_2$. Its role in deep-inelastic scattering
has been discussed extensively in the
literature.$^{11}$ Sometime ago, Burkhardt and
Cottingham proposed
a sum rule for $G_2$,$^{11}$
$$     \int^\infty_{Q^2/2} d\nu G_2(\nu, Q^2) = 0.   \eqno(10) $$
This sum rule was derived from the so-called
super-convergence condition and is also a
statement about rotational invariance.$^{13}$
Validity of the sum rule
has also been discussed in various places.$^{12,14,15}$
In the following discussion, I assume the sum rule is correct.
The elastic contribution to $G_2$ is simple to calculate,
$$     G_2^{\rm el} = - {M^2\over 2}F_2(F_1 + F_2)
              \delta(2\nu - Q^2).    \eqno(11) $$
Thus, eq. (10) can be written in another form,
$$     \int^\infty_{\nu_{\rm in}} d\nu G_2(\nu, Q^2) = {M^2\over 4}
     F_2(F_1+F_2) \eqno(12)  $$
where $\nu_{\rm in}$ is the inelastic threshold.

In a recent paper, Soffer and Teryaev$^{16}$ have
proposed that strong $Q^2$ dependence of
$\bar I_1(Q^2)$ can be understood from the
contamination of $G_2$ in
longitudinally-polarized photon scattering. Below, I re-examine
this suggestion.

Since $G_1$ and $G_2$ are invariant structure functions,
they both contribute to the helicity amplitudes
of longitudinally and transversely polarized
virtual-photon scattering.
For the longitudinally-polarized scattering,
$$      \sigma_{1\over 2} -  \sigma_{3\over 2}
          \sim G_1 - {Q^2\over \nu} G_2(\nu, Q^2),  \eqno(13)  $$
Thus it is natural to introduce an interpolating
sum rule,
$$ \eqalign{         K_1(Q^2) & = \int^{\infty}_{Q^2/2}
              {d\nu \over \nu} \big[G_1(\nu, Q^2)
           - {Q^2\over \nu} G_2\big],         \cr
             & = {2M^2\over Q^2}\int^1_0 \Big[
              g_1(x, Q^2) - {4M^2x^2\over Q^2}g_2(x, Q^2)\Big]dx.}
     \eqno(14) $$
Clearly, at high $Q^2$, we have,
$$        K_1(Q^2) \rightarrow I_1(Q^2).   \eqno(15) $$
However, at low-$Q^2$, the elastic part of
$G_2$ also contributes,
$$        K_1(Q^2) = {M^2\over Q^2}(F_1 + F_2)^2 -{\kappa^2\over 4}.
     \eqno(16) $$
Validity of the above result can be understood from the
fact that $(F_1+F_2)^2$
represents the amplitude of two helicity flips in
Compton scattering. It can be shown, however, that the
inelastic contribution to $G_2$ enters $K_1(Q^2)$
only at the order of $Q^2$. Thus if one is interested in
the subtracted version of the sum rule, we have
$$       \bar A_1(Q^2) \sim  \bar I_1(Q^2),      \eqno(17)    $$
at both high and low $Q^2$ limits, where the bar quantities
have the elastic contribution subtracted.

For scattering with transversely polarized nucleon
targets, the asymmetry is related to the interference
between the transversely and longitudinally polarized photons.
I define for this case
an interpolating sum rule,
$$ \eqalign{         K_2(Q^2) & = \int^{\infty}_{Q^2/2}
              {d\nu \over \nu} [G_1(\nu, Q^2)
           + \nu G_2(\nu Q^2)]         \cr
             & = {2M^2\over Q^2}\int^1_0
         [g_1(x, Q^2) + g_2(x, Q^2)]dx. }\eqno(18) $$
Clearly, because of the Burkhardt-Cottingham sum rule,
$$   K_2(Q^2) = I_1(Q^2), \eqno(19)  $$
at all $Q^2$. On the other hand, if
subtracting off the elastic contribution, I have at low $Q^2$,
$$   \bar K_2(0)  = \kappa /4   \eqno(20)  $$
where the contribution of the inelastic part of $G_2$
has been included from eq. (12).
At high $Q^2$, we still have,
$$       \bar K_2(Q^2) \rightarrow I_1(Q^2).  \eqno(21)         $$
Eq. (20) was obtained by Soffer and Teryeav and
was proposed as a solution to the rapid $Q^2$
variation of $\bar I(Q^2)$. It must be
emphasized, however, that this combination of $G_1$ and $G_2$
appears only in the transverse and longitudinal
interference of virtual-photon scattering.
Besides, the slow $Q^2$ variation of $\bar K_2(Q^2)$
covers up the rapid variation of $\bar I_1(Q^2)$
due to the nucleon resonances at low energy.

Thus, it appears that one can define a number of
interpolating sum rules,  $I_1(Q^2)$ and $K_{1,2}(Q^2)$
with and without the elastic contribution, to
interpolate the EMC data and the real photon point.
However, the most interesting
sum rule is $I_1(Q^2)$ with the elastic contribution
included, for which there may exists
a region of $Q^2$ where both resonance physics
and deep-inelastic physics are correct. If so, we can
extract interesting higher
twist matrix elements from the sum rule.
\vfill
\eject

\centerline{\bf REFERENCES}
\bigskip
\item{1.}V. Burkert et al., CEBAF Proposal 91-023, 1991;
S. E. Kuhn et al., CEBAF Proposal, 1993.
\medskip
\item{2.}J. Ashman et al., {\it Nucl. Phys.} {\bf B328} (1989) 1.
\medskip
\item{3.}S. D. Drell and A. C. Hearn, Phys. Rev. Lett.
{\bf 16} (1966) 908; S. B. Gerasimov, Sov. J. Nucl. Phys.
{\bf 2} (1966) 430.
\medskip
\item{4.}M. Anselmino, B. L. Ioffe, and E. Leader,
Sov. J. Nucl. Phys. {\bf 49} (1989) 136.
\medskip
\item{5.}M. J. Alguard el al., {\it Phys. Rev. Lett.}, {\bf 37} (1976)
1258.
\medskip
\item{6.}A. De Rujula, H. Georgi, and H. D. Politzer,
{\it Ann. Phys.} {\bf 103} (1977) 315.
\medskip
\item{7.}J. Ellis, Invited talk at Neutrino 79, preprint TH. 2701-CERN,
1979.
\medskip
\item{8.}V. D. Burkert and B. L. Ioffe, Phys. Lett. {\bf B296} (1992) 223.
\medskip
\item{9.}X. Ji and P. Unrau, to be published.
\medskip
\item{10.}E. V. Shuryak and A. I. Vainshtein, Nucl. Phys. {\bf B201}
(1982) 141; Y. Balitsky, V. M. Braun, and A. B. Kolesnichenko,
JETP Lett. {\bf 50} (1989) 54.
\medskip
\item{11.}See discussions and references in R. L. Jaffe and X. Ji,
Phys. Rev. {\bf D43} (1991) 724.
\medskip
\item{12.}H. Burkhardt and W. N. Cottingham, Ann. Phys.
(NY) 56, 453 (1970).
\medskip
\item{13.}R. P. Feynman, {\it Photon-Hadron Interactions},
Benjamin, New York, 1972.
\medskip
\item{14.}R. L. Heiman, {\it Nucl. Phys.} {\bf B64} (1973) 429.
\medskip
\item{15.}R. L. Jaffe, Commun. Nucl. Part. Phys. {\bf 14} (1990) 239.
\medskip
\item{16.}J. Soffer and O. Teryaev, Marseille preprint, 1992.
\medskip

\par
\vfill
\end